\documentclass{svproc}
\usepackage{graphicx}
\usepackage{listings}

\begin{document}

\title{A Case Study on Record Matching of Individuals in Historical Archives of Indigenous Databases}

\titlerunning{Historical Record Matching in Indigenous Databases}

\author{Matthew Currie \and Ramon Lawrence}

\authorrunning{Currie, Lawrence} 

\tocauthor{Matthew Currie, Ramon Lawrence}

\index{Currie, M.}
\index{Lawrence, R.}

\institute{
Department of Computer Science, University of British Columbia \\ Kelowna, BC, Canada, V1V 2Z3\\
\email{mfsc@student.ubc.ca, ramon.lawrence@ubc.ca}}

\maketitle

\begin{abstract}
Digitization of historical records has produced a significant amount of data for analysis and interpretation. A critical challenge is the ability to relate historical information across different archives to allow for the data to be framed in the appropriate historical context. This paper presents a real-world case study on historical information integration and record matching with the goal to improve the historical value of archives containing data in the period 1800 to 1920.  The archives contain unique information about Métis and Indigenous people in Canada and interactions with European settlers. The archives contain thousands of records that have increased relevance when relationships and interconnections are discovered. The contribution is a record linking approach suitable for historical archives and an evaluation of its effectiveness. Experimental results demonstrate potential for discovering historical linkage with high precision enabling new historical discoveries.
\end{abstract}

\noindent {\bf Keywords:} record linkage, record matching, historical, database, archive, Indigenous

\section{Introduction}

Digitized historical data allows historians to search, analyze, and relate data across a diversity of data sources and time periods for a variety of applications. As paper-based historical records are converted into electronic form and loaded into data archives, preserving data quality and integrity is critical. During the data loading process, records are preserved for historical accuracy and determining interconnections between data already in the archive is a challenging task. It is the interconnections between the data records that have significant value as only when individual records are connected is it possible to have a more detailed perspective on individuals through time. Interconnecting records in diverse archives requires techniques for record matching and linkage \cite{christen2011survey,winkler14} that have been developed for database integration.

Record matching in the historical context is challenged by both the imprecision of the data recording and transcription and the limited amount of data available. A common record matching application is identifying records that belong to the same person or customer in one or more databases \cite{christen2011survey}. Published approaches and commercial systems achieve high matching accuracy by exploiting detailed information in the records that typically have fields such as names, addresses, phone numbers, emails, and identifying information like national identifiers (e.g. social security number). In the historical context, the data available is limited to names and a date when the recorded event occurred. This limits the ability to perform accurate matching. Another key goal is to achieve high precision with few false positives so that people in historical records are only matched correctly. In many applications, it is preferable to have no match rather that suggesting incorrect linkages.

This work presents a case study on record matching for historical data sets present in the Digital Archive Database Project (DADP) that contains over 165,000 records on Métis, Indigenous people, and European settlers in Canada in the period 1800 to 1920. This unique data set has been created by the digitization and integration of numerous independent data sources. Historians require record linkage between the data sources to identify when a person is involved in multiple historical events. 

The contributions of this work are:

\begin{itemize}
    \item An analysis of specific requirements for historical record linkage in this context.
    
    \item An algorithm for historical record linkage that has high precision even when data records only contain names, dates, and approximate locations.
    
    \item A performance evaluation on the record linkage approach for a large, real-world historical data set.
    
\end{itemize}

The paper begins with a background on the DADP historical archive, the record linkage requirements, and a comparison with prior work on record linkage. A description of the record linkage approach is in Section 3. Section 4 discusses experimental results. The paper closes with future work and conclusions.

\section{Background}

\begin{figure}[!htb]
	\centering
	\includegraphics[width=3.4in]{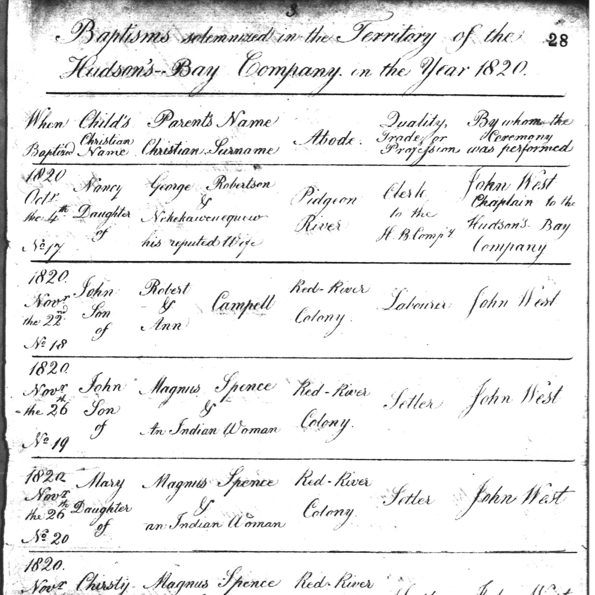}
	\caption{Example Historical Document}
	\label{archivepage}
\end{figure}

The Digital Archive Database Project (https://dadp.ok.ubc.ca) is a historical archive that aggregates a variety of historical documents in the period 1800 to 1920 from a diversity of sources in Canada. Individual data sources include information on sacramental records (birth, marriage, death) from a variety of missions and churches, census information, fur trade records, and voyageur contracts. Each data source is integrated into the archive by transcribing images of original source documents and annotating and standardizing information such as surnames (last names) and locations. An example image of a historical record is in Figure \ref{archivepage}.

The original source document image along with the transcribed information and metadata are stored in the archive.  Historians analyzed each record extracting fields such as name, location, and relationships between individuals in the record. Although the annotation process preserves accuracy and performs standardization, there remain issues with data quality and missing information. When the data is loaded into a database, each person in a historical record is added to the archive. For example, in a sacramental marriage record, numerous individuals will be recorded by name including the husband, wife, groom's parents, wife's parents, and witnesses. A particular individual may be referenced in numerous records at different times, and determining if a name in two or more records is the same individual is challenging even for historians with additional contextual information.

Matching individuals in the data archives by name and other metadata allows for a more complete historical record and the ability to follow individuals through time. This process improves the data quality and user experience and allows relationships to be made between records. These linkages lead to new historical knowledge and insights, but are very tedious to perform manually. DADP had no prior record linkage process prior to this case study, and the approaches used have a significant impact on the archive quality. 

Record matching techniques are well-known and often applied to match customers and people across databases. If the matching process is identifying and eliminating similar records in the same database, it is called deduplication. A survey of record matching and deduplication \cite{christen2011survey} details many approaches \cite{christen12book,winkler14}. The main difference with historical data sets is the limited amount of information for matching, typically only names and dates. This limits the applicability and accuracy of matching approaches.

Record matching for historical databases is comparatively less studied. Probabilistic approaches \cite{abramitzky19,abramitzky20} have been used for matching census data. This approach uses field similarity scores and the Expectation-Maximization algorithm to determine a likelihood that two historical records match. On aggregate over collections of census records, the algorithm produces matches consistent with manually matching census records suitable for aggregate, historical analysis of trends. Matching records in census applications has also been described in \cite{ruggles18}.

The focus of this particular work is matching individuals across diverse historical archives beyond only census data. Further, the goal is to be able to identify and track the historical events and relationships of individuals over time and space rather than performing large-scale aggregate analysis (e.g. how population and employment trends change over time). Due to this focus on individuals and historical accuracy, high precision is required for matching. Application of the Expectation-Maximization algorithm on the census data sets in DADP was successful in generating matches, but the precision was lower than required. Results for the sacramental data sets were unsatisfactory. Probabilistic record matching was not viable for these data sets as most records had only a name and limited other information. Census records often contain locations, ages, and birth dates allowing better matching opportunities. Statistical methods commonly used in matching census records were attempted but deemed ineffective largely due to sacramental data lacking the stronger identifying data available in census records such as birthplace and birthdate. 

Each sacramental record in DADP records a specific event in time at a location with participating individuals. For example, a marriage will involve a husband and wife, family members, and witnesses. The record will have a date and location and names of the individuals involved. A given person may be present in multiple records by name only. Consider a father with several daughters who get married. The father will be named in multiple marriage records as the ``father of the bride". The ability to use a relationship role in addition to a name is a unique feature of the matching required for the data archive and a key feature in the proposed approach. The fields utilized in matching individuals are first name, last name, location of the event (e.g. the church married/baptized in), approximate date of the event, the person’s role in the event, and other people also recorded in the event. 

In summary, record matching for historical data sets provides the opportunity to improve historical knowledge and context but requires specific approaches adapted to the limited field information available.

\section{Record Matching Approach}

The record matching approach produces matches with high precision on data sets that consist primarily of names, relationship roles, and event date and locations. Using relationship roles (e.g. father, mother, brother) improves accuracy, as it is quite common for two family members to appear in sacramental records together. 

\begin{figure}[!htb]
	\centering
	\includegraphics[width=3.4in]{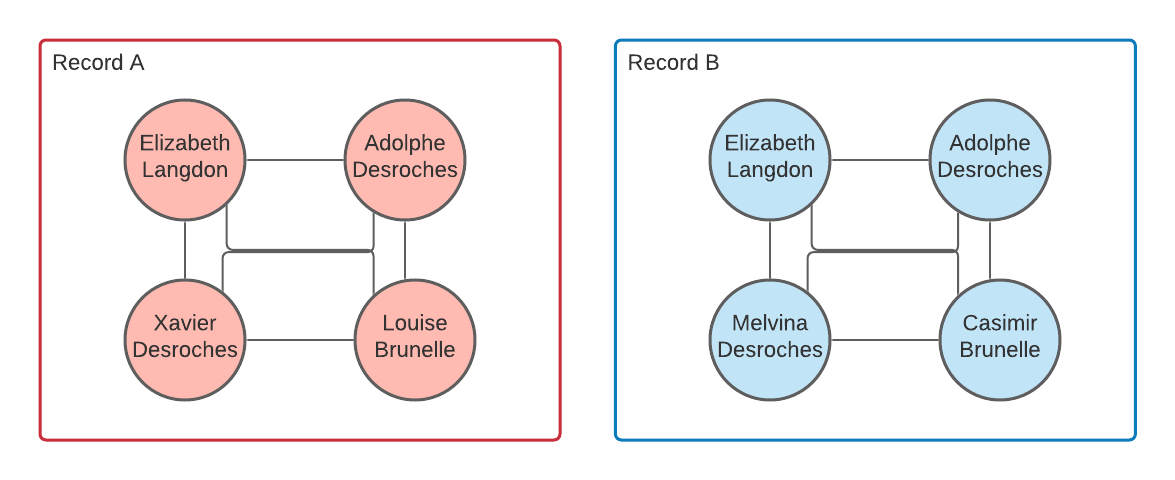}
	\caption{Using Relationships for Record Matching}
	\label{recconnections}
\end{figure}

Figure \ref{recconnections} illustrates an example of relationship roles that commonly occur in marriage records. The nodes on the graph represent a person contained within the marriage record while the edges indicate the connection between them. In the two records, Adolphe Desroches and Elizabeth Langdon appear in the marriage records and their roles in both indicate that they are a husband and wife and parents of the bride. Since familial dyads like this occur regularly in sacramental data, this information can be used as a strong indicator that the two records match.

The record matching approach consists of the following steps:

\begin{itemize}
    
\item Cleaning and standardization - names, locations, and dates are put into standardized formats

\item Indexing - records are grouped and indexed by key attributes (first and last name)

\item Matching and Grouping - records are matched based on names, locations, and relationship roles. Relationship roles are used to cluster records into groups.

\end{itemize}

An overview of the algorithm is in Figure \ref{alg}.

\begin{figure}[htbp]
\begin{lstlisting}[language=Java,basicstyle=\footnotesize\ttfamily]
Retrieve records from database archive
Perform data cleaning and standardization 
  on names, dates, and locations

Index and group records by name

for each index group:
    sort records in group by date
    
    for each record in group:
        if not in a record set:
            create new record set with record
            
        for each other record in group:
            if  name similarity above threshold and
                date records within range and
                location similarity match and
                relationships are consistent
            add matching record to record set
\end{lstlisting}
\caption{Pseudocode for Matching Implementation}
\label{alg}
\end{figure}

\subsection{Record Indexing}

The first step in the matching algorithm is indexing and grouping records for efficiency. With hundreds of thousands of records, it is inefficient to match every record with all others. Before matching, records are grouped and indexed by name similarity. Only records with names of sufficient similarity are compared. 

One of the strengths of the DADP archive is that records have first and last names standardized. This significantly eliminates issues with naming and indexing on name has high accuracy. The index used is on first and last name. Each index group is then sorted by record (event) date for efficient analysis. Sorting by date helps resolve issues such as comparing a father and son with a shared name. 

\subsection{Record Matching}

The matching process compares and groups records with a common index into sets of matching records. The input to the process is the set of index groups. The output, for each index group, are sets of matching records. If the algorithm determines that two or more records match, they are in the same set. If a given individual appears in multiple records, all of those records would be in the set. An advantage of this set-based approach is that record matching is improved as more records, with additional information, are added to the set.  It is possible that by using a  set of matching records that this set can match two records that do not meet all the matching criteria in isolation but do meet all the criteria when considering the other records in the set. This can happen because one record can act as a link between two records that would not have otherwise matched. For example, consider three records A, B, and C, such that A matches with B, B matches with C, but C does not match with A. The fact that B matches with both A and C will result in A and C being in the same set of matching records despite the fact that they would not match if B was not present. 

\subsubsection{Name Matching}

Names are the most critical components in record matching, especially in sacramental data. First name and last name are the closest attributes a person has to a unique identifier. The indexing step grouped records by name. The matching step performs similarity matching using various string comparison algorithms such as Levenshtein edit distance or Jaro-Winkler distance.

\subsubsection{Location Matching}

Locations, like names, are prone to some differences in how they are recorded. A church, for instance, can be referred to by different names such as the ``Rapids Church" also being referred to as the ``Grand Rapids Church". Location comparisons may use a combination of substring comparisons, edit distance, and Damerau-Levenshtein distance.

Unlike census data that contains a person's birthplace, sacramental data contains the location where an event occurred. Multiple records with similar locations are used to differentiate people with the same name but different locations where present. In combination with dates, it is possible to know accurately if two people are different if they are not able to travel between locations in a given time.

\subsubsection{Time Matching}

Although age or birth date are not commonly available in sacramental data, the time a record event is recorded is valuable information for matching.  For example, the name ``John Setter" appears in multiple historical records over time, specifically in several Rapids Church baptismal records. The dates the records were recorded can act as an indication for whether two records are matching. Since the records occur in 1847, 1848, and 1870, it is presumably more likely that the two records that are recorded within a one-year time span of one another are a match more than a record twenty years later. The matching algorithm is configurable to determine the time span between records that are still considered a match. In matching the sacramental data in the DADP, two records meet the matching criteria if they occur within a five-year time span of one another. This number was chosen through testing multiple values to determine the value yielding the most quality matches.  Note that matching records in sets helps with the time matching criteria as two records that do not meet the time span matching criteria in isolation can still be placed into the same set of matches through one or more linking records.

\subsubsection{Relationship Connections}

Using relationship connections is n unique contribution that improves accuracy. Utilizing any relational data available in the records is key to improving the quality of matches for sacramental data. In the DADP, familial dyads like a husband and wife frequently occur together across several historical records and are an important way that people are distinguishable from one another. The easiest way to compare relations is to treat all the people that occur within a record as nodes on a graph. A set intersection on the nodes of each graph reveals if the two records contain more than one probable match. Since false matches do not normally share more than one node on a graph, it is safe to assume that a familial relationship is present when a set intersection yields two or more commonalities. 

\begin{figure}[!htb]
	\centering
	\includegraphics[width=3.4in]{img/recconnections.png}
	\includegraphics[width=3.4in]{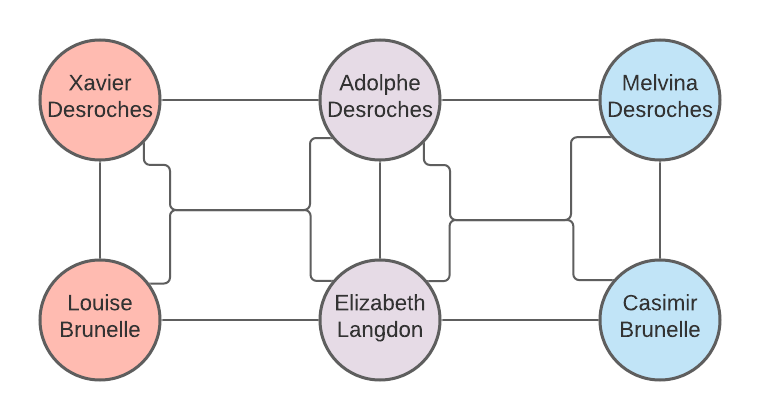}
	\caption{Using Relationships for Record Matching}
	\label{recrelationship}
\end{figure}

Figure \ref{recrelationship} illustrates how two source records are matched based on their common nodes. When a set intersection of Record A and Record B is performed, Adolphe Desroches and his wife, Elizabeth Langdon, are shown to appear in the two records together. Performing a set intersection on the indices of records will often yield person records that have a kind of familial relation. In the case illustrated with Adolphe and Elizabeth, this relation is a husband and wife.

Once it is determined that there are at least two probable matches within the pair of records, it is important to ensure that the relations present are valid. A common occurrence in the DADP marriage and baptismal records was a father and son with a common name being incorrectly matched due to their common relations. These erroneous matches were avoided by comparing the roles of a person in one record to another. Someone with the role of a husband record, for instance, should never match with someone that has a father role in a record that appears just a few years later. The same can be applied for death records as someone buried in 1870 should never match with anyone after 1870. Each database has specific validation rules that can be encoded to remove impossible matches based on relationship context and time.

\section{Case Study and Experimental Evaluation}

\begin{figure}[!htb]
	\centering
	\includegraphics[width=3.4in]{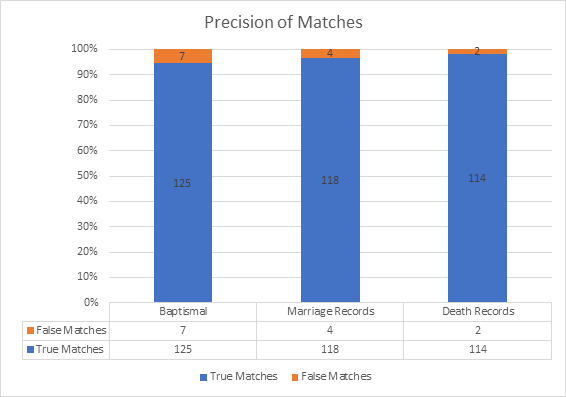}
	\caption{Matching Results}
	\label{results}
\end{figure}

The algorithm was tested on three data sources in the sacramental database of the DADP. Due to the sets of true matches and true nonmatches being an unknown, using common metrics for measuring accuracy such as F-score and recall was not possible. Therefore, it was necessary to use alternative methods for determining the effectiveness of the algorithm. A portion of the results for each table was manually verified for whether a set of records were a true match or a false match. The performance results are in Figure \ref{results}.

The matching precision is very high (over 95\%) and suitable for production use in DADP. The most common error made by the algorithm was erroneously matching family members, specifically fathers and their sons given that they share a name. Using relationship roles significantly reduces these errors, but they are still present. Even historians have difficulty in disambiguating references by the information in the records without having additional historical context.

The strengths of this approach to record matching is that it is highly flexible and can be made to work with a wide variety of data.  Figure 5 shows the results of applying the algorithm to three distinct data record types: baptismal, marriage, and death. These records have different formats and data fields. Using the relationship roles results in higher accuracy and exploits the unique information available in sacramental records.

An area for improvement is handling archives where many individuals share a common name. Additional information and rules on relationship roles may further help resolve differences in individuals. For example, three records of a Joseph Allery all originate from the same marriage record but each with different roles: first witness, husband’s father, and husband. Some measure must be taken in order to prevent matching the father and witness with the Joseph Allery being married, as they are clearly not referring to the same person despite having all of the common details. One suggestion for this scenario is to create a hierarchy of roles that are most likely to least likely to have true matches in the data. In addition, the matching algorithm for marriage records should have additional checks being made such that someone who is getting married is never matched with someone who is a father or mother. Ultimately, these issues will depend on the kinds of data that is being matched and solutions will have to be tailored where appropriate. Overall, the case study demonstrates matches produced that improve the quality and historical value of the DADP for historians.

\section{Conclusions and Future Work}

Historical record matching and deduplication presents unique challenges compared to matching data sets common in typical databases. The record matching algorithm must handle the limited number of data fields, often only names, and still generate matches with high precision. This work presented a real-world case study that extended current matching approaches to include data on relationship roles and family relationships to improve the matching accuracy. Using rules based on these relationships, times, and locations, 95\% of matches generated were true matches. These matches were used to improve the quality of the DADP archive and allow historians to visualize an individual's historical events and relationships over time and space. The approach and technique of using relationship roles to improve matching accuracy is generalizable to many other historical data sets and applications. Future work will be to produce more results on the precision and recall of the algorithm on the data set, which requires historians label the data set, and explore matching techniques to reduce the manual effort required. 

\section{Acknowledgment}

The authors would like to thank Aboriginal Affairs and Northern Development Canada for supporting this research.

\bibliographystyle{spmpsci}
\bibliography{refs}

\end{document}